\documentclass[]{raa}
\usepackage{graphicx,times}
\usepackage{natbib}
\usepackage{amssymb,amsmath}
\bibpunct{(}{)}{;}{a}{}{,}

\renewcommand{\vec}[1]{\mbox{\boldmath $#1$}}

\def\gsim{\lower.4ex\hbox{$\;\buildrel >\over{\scriptstyle\sim}\;$}}
\def\lsim{\lower.4ex\hbox{$\;\buildrel <\over{\scriptstyle\sim}\;$}}

\usepackage[a4paper=true,dvipdfm=true,pagebackref=true]{hyperref}
\hypersetup{pdftitle = The title of my PDF, pdfauthor = My name, pdfsubject= The subject}
\hypersetup{colorlinks = true, linkcolor = green, anchorcolor = red, citecolor = blue, filecolor = red, pagecolor = red, urlcolor = red}

\begin{document}

\title{
Dynamo Saturation in Rapidly Rotating Solar-Type Stars
}

\volnopage{ {\bf 2015} Vol.\ {\bf X} No. {\bf XX}, 000--000}
   \setcounter{page}{1}

\author{
L.\,L.~Kitchatinov\inst{1,2}, S.\,V.~Olemskoy\inst{1}
   }

   \institute{Institute for Solar-Terrestrial Physics, Lermontov Str. 126A, 664033, Irkutsk, Russia; {\it kit@iszf.irk.ru}
        \and
   Pulkovo Astronomical Observatory, St. Petersburg, 196140, Russia
   \\
   {\small Received 2015 XXXX; accepted XXXX}
}

\abstract{The magnetic activity of solar-type stars generally increases with stellar rotation rate. The increase, however, saturates for fast rotation. The Babcock-Leighton mechanism of stellar dynamos saturates as well when the mean tilt-angle of active regions approaches ninety degrees. Saturation of magnetic activity may be a consequence of this property of the Babcock-Leighton mechanism. Stellar dynamo models with a tilt-angle proportional to the rotation rate are constructed to probe this idea. Two versions of the model - treating the tilt-angles globally and using Joy's law for its latitude dependence - are considered. Both models show a saturation of dynamo-generated magnetic flux at high rotation rates. The model with latitude-dependent tilt-angles also shows a change in dynamo regime in the saturation region. The new regime combines a cyclic dynamo at low latitudes with an (almost) steady polar dynamo.
}

   \authorrunning{L.\,L.~Kitchatinov \& S.\,V.~Olemskoy}            %author_head in even pages
   \titlerunning{Dynamo Saturation in Rapidly Rotating Stars}  % title_head in odd pages
   \maketitle

%%%%%%%%%%%%%%%%%%%%%%%%%%%%%%%%%%%%%%%%%%%%%%%%%%%%%%%%%%%%%%%%%%%%%%%
\section{Introduction}
%%%%%%%%%%%%%%%%%%%%%%%%%%%%%%%%%%%%%%%%%%%%%%%%%%%%%%%%%%%%%%%%%%%%%%%
Sun-like stars are expected to host hydromagnetic dynamos generating magnetic fields in their external convection zones. This expectation is supported by the observed correlation of stellar magnetic activity with rotation rate. The physical reason for the correlation is the key role of rotation in dynamos  \citep{P79}. On the Sun, chromospheric Ca\,II emission and coronal X-ray emission are correlated with magnetic fields. The relative Ca\,II and X-ray luminosities are, therefore, accepted as measures of stellar magnetic activity. Chromospheric \citep{Nea84,SB99} and coronal \citep{V84,VW87,Pea03,Wea11} activities generally increase with stellar rotation rate.

The increase in X-ray emission, however, saturates for Rossby numbers $Ro \lsim 0.1$; $Ro = P_\mathrm{rot}/\tau_\mathrm{c}$ is the ratio of the rotation period to convective turnover time. Coronal activity stops increasing with sufficiently rapid rotation \citep{Wea11}. Saturation in chromospheric activity, though less pronounced, can also be seen in Fig.\,6 of \citet{Nea84}. The observed saturation is an important clue for the dynamo theory of stellar activity. It indicates that dynamos do not produce stronger magnetic fields when the rotation rate of a star increases beyond a certain limit. A conventional explanation for dynamo-saturation is, however, still lacking.

Literature on stellar dynamos is vast. \citet{Bea94}, \citet{Jea10} and \citet{Iea11} constructed dynamo models for stars with different rotational velocities but did not consider the saturation problem. \citet{Kea14} discussed the possibility that a change in the direction of magnetic buoyancy from radial to parallel to the rotation axis as the rotation rate increases \citep{CG87,SS92,W11} may cause the dynamo-saturation. They concluded that though the change in buoyant rise direction reduces the dynamo's efficiency, this effect does not suffice to explain the dynamo saturation. More recently, \citet{BT15} suggested that an increase in differential rotation with angular velocity may cause the saturation. The life-time of convective eddies in rapidly rotating stars may be reduced due to distraction of the eddies by rotational shear thus leaving less time for Coriolis force to affect the eddies and reduce dynamo efficiency. However, neither observations \citep{Bea05,C07} nor modeling \citep{KO12} of stellar differential rotation show its considerable increase with rotation rate.

This paper concerns the possibility that dynamo saturation in rapid rotators is caused by saturation in the Babcock-Leighton (BL) mechanism of poloidal field generation. This mechanism was introduced by \citet{B61} and first used in a dynamo model by \citet{L69}. It is related to the observed average tilt of bipolar groups of sunspots relative to the local parallel of latitude \citep{Hea19,H96}. Due to the tilt, magnetic fields of solar active regions have on average a finite poloidal component, which contributes to the global poloidal field upon the sunspot groups decay. There is growing evidence for the operation of the BL
mechanism on the Sun \citep{E04,Dea10,KO11,K14}. BL-type dynamo model of \citet{Jea13} reproduces the basic correlations observed in solar activity. The BL mechanism can be expected to participate in stellar dynamos as well.

The contribution of the BL mechanism to the global poloidal field formation is proportional to sine of the (averaged) tilt-angle. The tilt is believed to result from the Coriolis force and therefore can be expected to increase with rotation rate. The BL mechanism obviously saturates when the tilt angle approaches $\pi /2$. The characteristic value of the tilt angle for the Sun is about $5^\circ$ \citep{H96}. Assuming the angle to be proportional to the rotation rate, the tilt angle can reach $90^\circ$ in a (solar mass) star rotating about 20 times faster than the Sun. This is roughly where the saturation in coronal activity of solar-type stars is observed \citep{Wea11}.

To assess this idea quantitatively, we use stellar dynamo models, which only differ from a solar model by a modification of the $\alpha$-effect \citep[cf., e.g.,][]{KR80} of poloidal field regeneration. Two versions of the $\alpha$-effect dependence on rotation rate are considered. In one of them, the dependence is treated globally by multiplying the $\alpha$-effect by the factor of $\sin (\overline{\alpha}_\odot P_\odot / P_\mathrm{rot})$ where $\overline{\alpha}_\odot$ and $P_\odot$ are the averaged tilt-angle and rotation period of the Sun, respectively, and $P_\mathrm{rot}$ is the rotation period of a star. In the other model, a similar procedure is applied to the Joy's law of the latitude-dependent tilt-angle. The second model shows a wider variety of dynamo behaviours but both show saturation of the generated magnetic flux for rapid rotation. The models design is described in Section \ref{models}. The results of numerical computations for these models are discussed in Section \ref{results}. Section \ref{disc} summarises our main findings and concludes.
%%%%%%%%%%%%%%%%%%%%%%%%%%%%%%%%%%%%%%%%%%%%%%%%%%%%%%%%%%%%%%%%%%%%%%%
\section{Dynamo model}\label{models}
%%%%%%%%%%%%%%%%%%%%%%%%%%%%%%%%%%%%%%%%%%%%%%%%%%%%%%%%%%%%%%%%%%%%%%%
The dynamo model of this paper differs little from those used before and, therefore, it is described here only briefly. The only difference with \citet{OK13} is in formulation of the $\alpha$-effect. Other model ingredients like differential rotation, meridional flow and eddy transport coefficients remain unchanged. They may also depend on rotation rate but we avoid modifying anything else but the $\alpha$-effect to see the consequence of saturation in the BL mechanism alone which is not disguised by other modifications.
%%%%%%%%%%%%%%%%%%%%%%%%%%%%%%%%%%%%%%%%%%%%%%%%%%%%%%%%%%%%%%%%%%%%%%%
\subsection{Dynamo Equations}
%%%%%%%%%%%%%%%%%%%%%%%%%%%%%%%%%%%%%%%%%%%%%%%%%%%%%%%%%%%%%%%%%%%%%%%
We consider the axisymmetric magnetic field $\vec B$ in a spherical shell modeling the external convection zone of a star,
\begin{equation}
    \vec{B} = \vec{e}_\phi B + \vec{\nabla}\times\left( \vec{e}_\phi\frac{A}{r\sin\theta}\right) ,
    \label{1}
\end{equation}
where standard spherical coordinates $(r,\theta,\phi)$ are used, $\vec{e}_\phi$ is the azimuthal unit vector, $B$ is the toroidal magnetic field, and $A$ is the poloidal field potential. A similar expression for axisymmetric global flow $\vec V$ in a co-rotating frame can be written,
\begin{equation}
    \vec{V} = \vec{e}_\phi r\sin\theta\Delta\Omega f(r,\theta)
    + \rho^{-1}\vec{\nabla}\times\left( \vec{e}_\phi\frac{\psi}{r\sin\theta}\right),
    \label{2}
\end{equation}
where $\Delta\Omega$ is the characteristic value of the angular velocity variation within the convection zone, $f$ is the normalised differential rotation, $\rho$ is density, and $\psi$ is the stream-function of the meridional flow.

The dynamo equations are written with normalised variables. Distance is measured in units of stellar radius $R$, time - in the diffusive units of $R^2\eta_0^{-1}$, where $\eta_0$ is the characteristic value of the eddy diffusivity. The toroidal magnetic field is normalised to its \lq equipartition' strength $B_0$ for which nonlinear feedback on the $\alpha$-effect becomes essential. The poloidal field potential is normalised to $\alpha_0 B_0 R^3\eta_0^{-1}$, where $\alpha_0$ is the measure of the alpha-effect. Density is normalised to its value $\rho_0$ on the top boundary. The stream-function is normalised to $\rho_0R^2V_0^{-1}$, where $V_0$ is the amplitude of the surface meridional flow. The same notations are used for normalised variables as for their dimensional counterparts, except for the fractional radius $x = r/R$.

The normalised equation for the toroidal field of our dynamo model reads
\begin{eqnarray}
    \frac{\partial B}{\partial t} &=&
    \frac{\cal D}{x} \left(\frac{\partial f}{\partial x}\frac{\partial
    A}{\partial\theta} - \frac{\partial f}{\partial\theta}
    \frac{\partial A}{\partial x}\right)
    + \frac{R_\mathrm{m}}{x^2\rho}\frac{\partial\psi}{\partial x}
    \frac{\partial}{\partial\theta}\left(\frac{B}{\sin\theta}\right)
    - \frac{R_\mathrm{m}}{x\sin\theta}\frac{\partial\psi}{\partial\theta}
    \frac{\partial}{\partial x}\left(\frac{B}{\rho x}\right)
    \nonumber \\
    &+&  \frac{\eta}{x^2}\frac{\partial}{\partial\theta}\left(
    \frac{1}{\sin\theta}\frac{\partial(\sin\theta
    B)}{\partial\theta}\right)
    + \frac{1}{x}\frac{\partial}{\partial
    x}\left(\sqrt{\eta}\ \frac{\partial(\sqrt{\eta}\ xB)}
    {\partial x}\right) ,
    \label{3}
\end{eqnarray}
where
\begin{equation}
    {\cal D} = \frac{\alpha_0 \Delta\Omega R^3}{\eta_0^2}\
    \label{4}
\end{equation}
is the dynamo number and
\begin{equation}
    R_\mathrm{m} = \frac{V_0 R}{\eta_0}
    \label{5}
\end{equation}
is the magnetic Reynolds number for the meridional flow. The right-hand side of Eq.\,(\ref{3}) accounts for the toroidal field production by differential rotation, advection by the meridional flow, turbulent diffusion, and diamagnetic pumping due to the convective turbulence inhomogeneity \citep[see][hereafter KO12, for more details]{KO12_1}. The square root of $\eta$ appears in the last term to allow for diamagnetic pumping.

The poloidal field equation differs by representation of the $\alpha$-effect but is otherwise identical to the solar dynamo model of KO12,
\begin{eqnarray}
    \frac{\partial A}{\partial t} &=&
    x \sin\theta\ F(\theta,P_\mathrm{rot})
    \int\limits_{x_\mathrm{i}}^x
    \hat\alpha (x,x') B(x',\theta)\ \mathrm{d} x'
    + \frac{R_\mathrm{m}}{\rho x^2 \sin\theta}
    \left(\frac{\partial\psi}{\partial x}
    \frac{\partial A}{\partial\theta} -
    \frac{\partial\psi}{\partial\theta}
    \frac{\partial A}{\partial x}\right)
    \nonumber \\
    &+& \frac{\eta}{x^2}\sin\theta\frac{\partial}{\partial\theta}
    \left(\frac{1}{\sin\theta}\frac{\partial
    A}{\partial\theta}\right) + \sqrt{\eta}\frac{\partial}{\partial
    x} \left(\sqrt{\eta}\frac{\partial A}{\partial x}\right)  .
    \label{6}
\end{eqnarray}
The first term on the right-hand side of this equation stands for the non-local $\alpha$-effect of BL type. The kernel function $\hat\alpha(x,x')$ in this term differs considerably from zero only if $x'$ is close to the inner boundary $x_\mathrm{i}$ and simultaneously $x$ is close to the top boundary:
\begin{eqnarray}
    \hat\alpha (x,x') &=& \frac{\phi_\mathrm{b}(x')\phi_\alpha (x)} {1 + B^2(x',\theta)}
    ,
    \nonumber \\
    \phi_\mathrm{b}(x') &=& \frac{1}{2}\left( 1 -
    \mathrm{erf}\left( (x' - x_\mathrm{b})/h_\mathrm{b}\right)\right) ,\ \
    \phi_\alpha (x) = \frac{1}{2}\left( 1 +
    \mathrm{erf}\left( (x - x_\alpha)/h_\alpha\right)\right),
    \label{7}
\end{eqnarray}
where $B^2$ in the denominator represents the only nonlinearity of our model, which is the magnetic quenching of the $\alpha$-effect, and erf is the error function. The values $x_\mathrm{b} = x_\mathrm{i} + 2.5h_\mathrm{b}$ and $x_\alpha = 1 - 2.5h_\alpha$ ensure smoothness of the $\hat\alpha$-function; $h_\alpha = 0.02$ and $h_\mathrm{b} = 0.002$ in the computations of this paper. The $\alpha$-effect in our model, therefore, describes the poloidal field generation near the surface from the toroidal field of the thin near-bottom layer. The factor $x\sin\theta$ in the first term on the right-hand side of (\ref{6}) corresponds to the definition of the poloidal field potential $A$ by equation (\ref{1}). The function $F(\theta, P_\mathrm{rot})$ accounts for the dependence of the $\alpha$-effect on stellar rotation rate. This function will be defined in the next subsection.

The initial value problem for the equations (\ref{3}) and (\ref{6}) is solved numerically. The initial field is a mixture of dipolar and quadrupolar fields, i.e., the equatorial symmetry is not prescribed. The superconductor boundary condition is imposed at the bottom boundary and a pseudo-vacuum condition (zero toroidal and radial poloidal fields) at the top.
%%%%%%%%%%%%%%%%%%%%%%%%%%%%%%%%%%%%%%%%%%%%%%%%%%%%%%%%%%%%%%%%%%%%%%%
\subsection{Rotation Rate Dependence of the $\alpha$-Effect}\label{twomodels}
%%%%%%%%%%%%%%%%%%%%%%%%%%%%%%%%%%%%%%%%%%%%%%%%%%%%%%%%%%%%%%%%%%%%%%%
The poloidal part of the magnetic flux of a bipolar sunspot group is proportional to the sine of the group tilt-angle $\alpha$. We, therefore, assume the BL type $\alpha$-effect to be proportional to $\sin (\alpha )$ as well. The finite tilt in turn is believed to be caused by the Coriolis force action on the rising magnetic loops. The tilt-angle is therefore expected to increase with rotation rate of a star. The $\alpha$-effect changes little when the $\sin (\alpha )$ passes through a broad maximum near $\alpha \simeq 90^\circ$ with an increasing rotation rate. This saturation of the $\alpha$-effect may be the reason for the saturation in stellar magnetic activity.

The tilt angle in a solar mass star rotating with a period $P_\mathrm{rot}$ can be estimated as $\alpha = \alpha_\odot P_\odot /P_\mathrm{rot}$. Two models for the dependence of the $\alpha$-effect on the stellar rotation rate will be considered. In the first model, we use a \lq global' estimation of the tilt-angle to write
\begin{equation}
    F(\theta,P_\mathrm{rot}) = \cos\theta \sin^2\theta
    \frac{\sin(\overline{\alpha}_\odot P_\odot /P_\mathrm{rot})}{\sin\overline{\alpha}_\odot}
    \ \ \ -\ \mathrm{Model\ I} ,
    \label{8}
\end{equation}
where $\overline{\alpha}_\odot$ is the mean value of the tilt-angle for the Sun. The $\sin\overline{\alpha}_\odot$ was included as a denominator in this equation in order for the dynamo equations with $F$-function of Eq.\,(\ref{8}) to turn into our former solar model (KO12) at $P_\mathrm{rot} = P_\odot$. Then, the value of dynamo number ${\cal D} =3.2\times 10^4$ for which reasonable results for the Sun were obtained does not have to be changed. In the computations to follow $\overline{\alpha}_\odot = 5^\circ$ \citep{H96}.

In the second model, Joy's law for the latitude dependence of the tilt-angle \citep{Hea19} is allowed for. \citet{Dea10} found $\alpha_\odot = (0.26 \pm 0.05)\lambda$ and $\alpha_\odot = (0.28 \pm 0.06)\lambda$ from sunspot data of the Mount Wilson and Kodaikanal observatories, respectively, where $\lambda$ is latitude. \citet{I12} confirmed their results but found a somewhat larger proportionality factor of $0.38 \pm 0.03$ from the Pulkovo observatory database. \citet{SK12} infered a still steeper dependence of $\alpha_\odot = (0.56 \pm 0.01)\sin\lambda$ from {\sl MDI/SOHO} magnetograms. They also noted that $\sin\lambda$ is a natural choice for the fit function in Joy's law because the Coriolis force, presumably producing the tilts, varies as $\sin\lambda$ with latitude. The larger tilt-angles from magnetograms are probably explained by the contribution of plages \citep{Wea15}. As the BL type $\alpha$-effect is more related to sunspots of active regions than to smaller magnetic features of magnetograms, we employ Joy's law in the form $\alpha_\odot = 0.3\cos\theta$ to find
\begin{equation}
    F(\theta,P_\mathrm{rot}) = \sin\left(0.3\cos\theta\ P_\odot/ P_\mathrm{rot}\right)
    \ \ \ -\ \mathrm{Model\ II}
    \label{9}
\end{equation}
for the $F$-function in Eq.\,(\ref{6}). As computations of the solar dynamo with such a profile of the $\alpha$-effect have never been attempted before, we have to define the critical value of $\cal D$ (\ref{4}) for the onset of dynamo. The computed critical value for dipolar modes in the solar ($P_\mathrm{rot} = P_\odot$) model is ${\cal D}^\mathrm{d} = 6.63\times 10^4$ ($D^\mathrm{q} = 7.21\times 10^4$ for modes of quadrupolar parity). The solar dynamo is probably only slightly supercritical \citep[see discussion in][]{KKB14}. All computations in Model~II were performed with ${\cal D} = 7\times 10^4$, which is about 5\% supercritical for the solar case.

\begin{figure}
   \centering
   \includegraphics[width=10 truecm, angle=0]{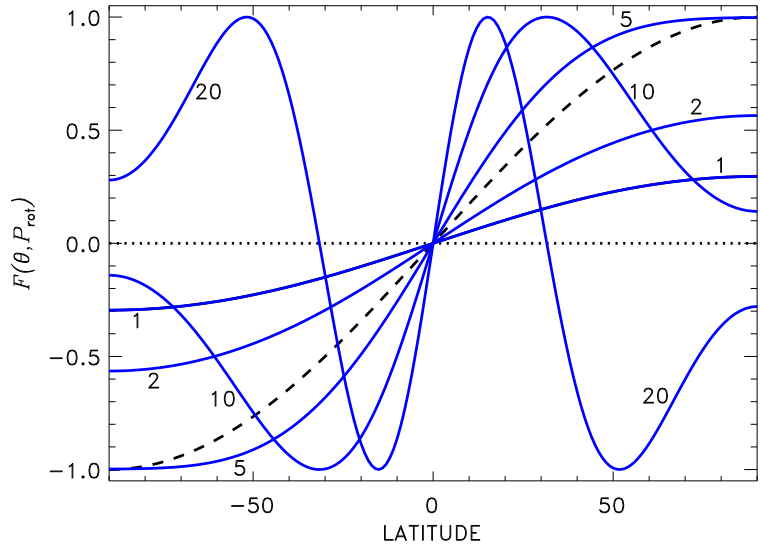}
   \caption{Latitudinal profiles of the $\alpha$-effect (\ref{9}) of Model II for different rotation
            rates. The lines are marked by the corresponding values of $P_\odot /P_\mathrm{rot}$. The often assumed $\cos\theta$-profile is shown for comparison by the dashed line.}
   \label{f1}
\end{figure}

Figure\,\ref{f1} shows latitudinal profiles of the $\alpha$-effect (\ref{9}) of Model~II for several rotation rates. The profile of $\cos\theta$ often assumed in dynamo models is shown for comparison. The $\alpha$-effect initially increases with rotation rate, but then saturates and even changes to a decrease at high latitudes for $P_\mathrm{rot} < P_\odot/5$. This is the reason for dynamo saturation in  subsequent computations. Note the sign reversals in the alpha profile for $P_\mathrm{rot} = P_\odot/20$. The change of sign is important for interpreting the results of dynamo simulations for rapid rotators.
%%%%%%%%%%%%%%%%%%%%%%%%%%%%%%%%%%%%%%%%%%%%%%%%%%%%%%%%%%%%%%%%%%%%%%%
\subsection{Other Model Ingredients}
%%%%%%%%%%%%%%%%%%%%%%%%%%%%%%%%%%%%%%%%%%%%%%%%%%%%%%%%%%%%%%%%%%%%%%%
The differential rotation, meridional flow and diffusivity profile in this paper are the same as in \citet{OK13}.

As mentioned in the Introduction, observations and theoretical modeling both suggest that differential rotation varies moderately with the rotation rate of a solar-type star of a given mass \citep{Bea05,C07,KO12}. The same differential rotation specified after the approximation of \citet{BKS00} for the helioseismological rotation law is used for all stars.

\begin{figure}
   \centering
   \includegraphics[width=10 truecm, angle=0]{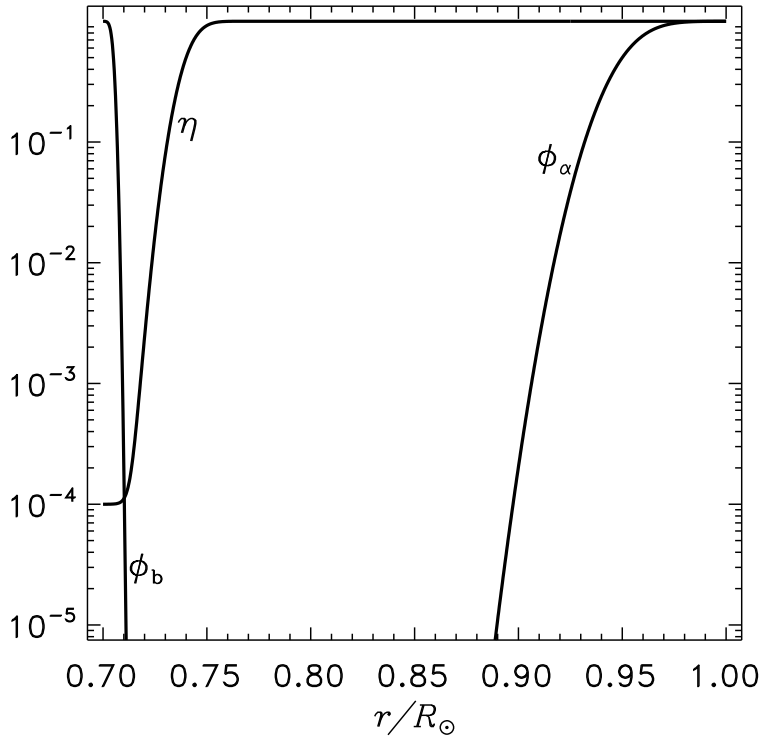}
   \caption{Profiles of diffusivity and the kernel functions (\ref{7}) of the non-local
            $\alpha$-effect of Eq.\,(\ref{6}) used in our dynamo models.}
   \label{f2}
\end{figure}

The eddy magnetic diffusivity varies little in the bulk of the solar convection zone but drops sharply with depth near its base. The diffusivity profile of our model mimics this behaviour:
\begin{equation}
    \eta (x) = \eta_\mathrm{in} + \frac{1}{2}(1 - \eta_\mathrm{in})
    \left( 1 + \mathrm{erf}\left(\frac{x -
    x_\eta}{h_\eta}\right)\right) .
    \label{10}
\end{equation}
Computations were performed for the following values of the parameters in this equation: $\eta_\mathrm{in} = 10^{-4}$, $x_\eta = 0.74$ and $h_\eta = 0.02$. Figure\,\ref{f2} shows the diffusivity profile for these parameter values together with the functions $\phi_\mathrm{b}$ and $\phi_\alpha$ of Eq.\,(\ref{7}) defining the alpha-effect. The diffusivity drops four orders in magnitude near the bottom boundary. A non-uniform numerical grid with very small grid spacing near the bottom was used to resolve this sharp variation.

Single-cell meridional circulation with poleward flow on the top and return equatorward flow near the bottom is prescribed \citep[see Fig.\,3 and Eq.\,(13) in][]{OK13}. The value of $R_\mathrm{m} = 10$ for the Reynolds number (\ref{5}) is used in the computations, which corresponds to the flow amplitude $V_0 \simeq 15$\,m\,s$^{-1}$ and background diffusivity $\eta_0 \simeq 10^9$\,m$^2$\,s$^{-1}$. With this relatively low Reynolds number, the meridional flow in the bulk of the convection zone is not significant. Only the flow near the bottom where diffusion is low (Fig.\,\ref{f2}) is important for magnetic field dynamics \citep{HKC14}. The same value of $\eta_0 \simeq 10^9$\,m$^2$\,s$^{-1}$ was used to convert dimensionless time into physical units.

As the BL $\alpha$-effect is related to surface active regions, we illustrate toroidal fields of dynamo models by butterfly diagrams for the same near-bottom toroidal flux,
\begin{equation}
    {\cal B}(\theta ) = \sin\theta \int\limits_{x_\mathrm{i}}^1
    \phi_\mathrm{b}(x) B(x,\theta)\ \mathrm{d} x ,
    \label{11}
\end{equation}
to which the $\alpha$-effect of Eqs.\,(\ref{6}) and (\ref{7}) is proportional. Note that the weight function $\phi_\mathrm{b}$ of Fig.\,\ref{f2} only differs considerably from zero in a thin near-bottom layer. Equation~(\ref{11}) therefore represents the near-bottom toroidal flux. The factor $\sin\theta$ in Eq.\,(\ref{11}) accounts for the dependence of the length of toroidal flux-tube on latitude. The probability of the production of spots is assumed to be proportional to this length. The unsigned total near-bottom flux,
\begin{equation}
    f_\mathrm{m} = \int\limits_{x_\mathrm{i}}^1\!\!\!\int\limits_0^\pi\! \sin\theta\,x\,\phi_\mathrm{b}(x)\,|B(x,\theta)|\,\mathrm{d}x\,\mathrm{d}\theta ,
    \label{12}
\end{equation}
is used as a proxy for the overall magnetic activity.

Dynamo computations started from a weak poloidal field with a mixed equatorial parity. Upon several diffusion times, initial field growth saturates. The results of the next section correspond to these saturated dynamo-regimes.
%%%%%%%%%%%%%%%%%%%%%%%%%%%%%%%%%%%%%%%%%%%%%%%%%%%%%%%%%%%%%%%%%%%%%%%
\section{Results and discussion}\label{results}
%%%%%%%%%%%%%%%%%%%%%%%%%%%%%%%%%%%%%%%%%%%%%%%%%%%%%%%%%%%%%%%%%%%%%%%
In all runs, the dynamo-fields eventually approached dipolar parity. This seems to be a common feature of dynamo models with relatively large diffusivity in the bulk of the convection zone \citep{CNC04,JCC07,HY10}. The global magnetic field of the Sun is also close to dipolar parity \citep{S88}. All the simulated dynamos are cyclic.
%%%%%%%%%%%%%%%%%%%%%%%%%%%%%%%%%%%%%%%%%%%%%%%%%%%%%%%%%%%%%%%%%%%%%%%
\subsection{Model I}
\begin{figure}
   \centering
   \includegraphics[width=10 truecm, angle=0]{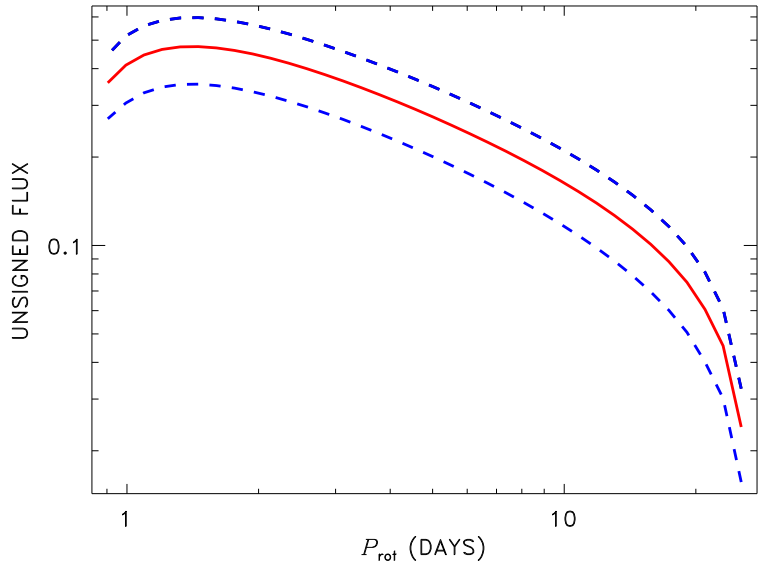}
   \caption{Dependence of unsigned flux (\ref{12}) on rotation period for Model~I.
            The dashed lines show maximum and minimum values of the flux in the dynamo-cycles. The full line shows the median value (half-sum of maximum and minimum).}
   \label{f3}
\end{figure}

%%%%%%%%%%%%%%%%%%%%%%%%%%%%%%%%%%%%%%%%%%%%%%%%%%%%%%%%%%%%%%%%%%%%%%%
Observational statistics of stellar activity usually do not cover epochs comparable to expected activity cycles. It is therefore not clear for which phase of the dynamo-cycles the fluxes (\ref{12}) should be estimated. Figure~\ref{f3} shows the maximum, minimum and median values of the fluxes in dependence on the rotation period.

The main feature of this plot is that the $f_\mathrm{m}$ stops increasing with rotation rate for small $P_\mathrm{rot}$. Observations of stellar activity show its saturation at Rossby numbers 10--20 times smaller than the solar value \citep[Fig.\,2 of][]{Wea11}. Stellar structure in our computations is not varied (because the Sun remains the only star for which the tilt-angles are measured). The Rossby number is, therefore, proportional to the rotation period. The position of the saturation region in Fig.\,\ref{f3} roughly agrees with observations. This figure also shows a decline in $f_\mathrm{m}$ for still shorter $P_\mathrm{rot}$. It is not clear whether the decline is real or not because observational statistics have considerable scatter and usually do not distinguish spectral types \citep[cf., however,][]{Pea03}.

Coronal activity data for moderately rotating stars ($Ro > 0.2$) are customarily fitted by the power law \citep{Pea81,Wea11},
\begin{equation}
    L_X \propto Ro^{-\gamma} .
    \label{13}
\end{equation}
The slope of the plot in Fig.\,\ref{f3} is not constant. However, it varies little in the region of $3 < P_\mathrm{rot} < 10$\,days. The slope $\delta$ defined in the same way as in \citet{Kea14},
\begin{equation}
    f_\mathrm{m}^2 \propto P_\mathrm{rot}^{-\delta} ,
    \label{14}
\end{equation}
is about $\delta \simeq 1.5$ in this region. Though our model is designed to qualitatively explain activity saturation in rapid rotators and not to achieve quantitative agreement with observations, a preliminary comparison is tempting. If $\gamma$ is taken to be equal to its canonical value of 2, then our model would mean that $L_X$ is proportional to $f_\mathrm{m}$ to the power of 2.7, which is somewhat larger than the power of 2 expected for $L_X$ being proportional to magnetic energy. The power index is still larger ($2\gamma/\delta \simeq 3.6$) for the value of $\gamma = 2.7$ of \citet{Wea11}.

Measurements of large-scale stellar magnetic fields are most relevant to predictions of dynamo models. The measurements are provided by the Zeeman-Doppler imaging (ZDI) technique \citep{DB97}. Though statistics of ZDI for main-sequence dwarfs is not vast (about 100 stars), the dependence of the large-scale fields on rotation rate has been estimated by \citet{Vea14}. Their results generally confirm the proportionality of X-ray luminosity to the square of unsigned magnetic flux $\Phi_V$ of large-scale fields. \citet{Vea14} also found the power law $\Phi_V \propto Ro^{-1.19\pm 0.14}$ for $Ro > 0.1$ and saturation of $\Phi_V$ at smaller $Ro$. We have to conclude that our dynamo model underestimates the power index $\delta$ of Eq.\,(\ref{14}).

\begin{figure}
   \centering
   \includegraphics[width=10 truecm, angle=0]{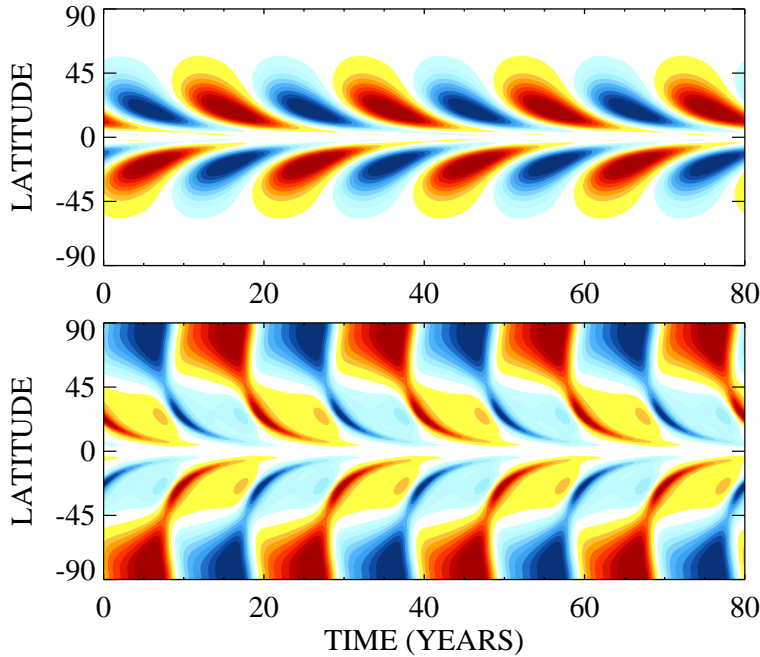}
   \caption{Time-latitude diagram of Model I for $P_\mathrm{rot} = 1$ day. {\sl Top panel:}
            near-bottom toroidal flux of Eq.\,(\ref{11}). {\sl Bottom panel:} radial magnetic field on the surface.
            }
   \label{f4}
\end{figure}

Figure~\ref{f4} shows the butterfly diagram for the rapid rotator with $P_\mathrm{rot} = 1$~day. It is quite similar to the solar model (KO12). Field patterns computed for different rotation rates are qualitatively similar and differ mainly in the field amplitude. This is because only the $\alpha$-effect in our model depends on rotation rate. The cycle period for Fig.\,\ref{1} is about 10 yrs. The cycle period changes little with rotation rate. This is because the meridional flow is not varied. The cycle period is controlled by the flow circulation time in advection-dominated dynamos \citep{C10,C11}.
%%%%%%%%%%%%%%%%%%%%%%%%%%%%%%%%%%%%%%%%%%%%%%%%%%%%%%%%%%%%%%%%%%%%%%%
\subsection{Model II}
\begin{figure}
   \centering
   \includegraphics[width=10 truecm, angle=0]{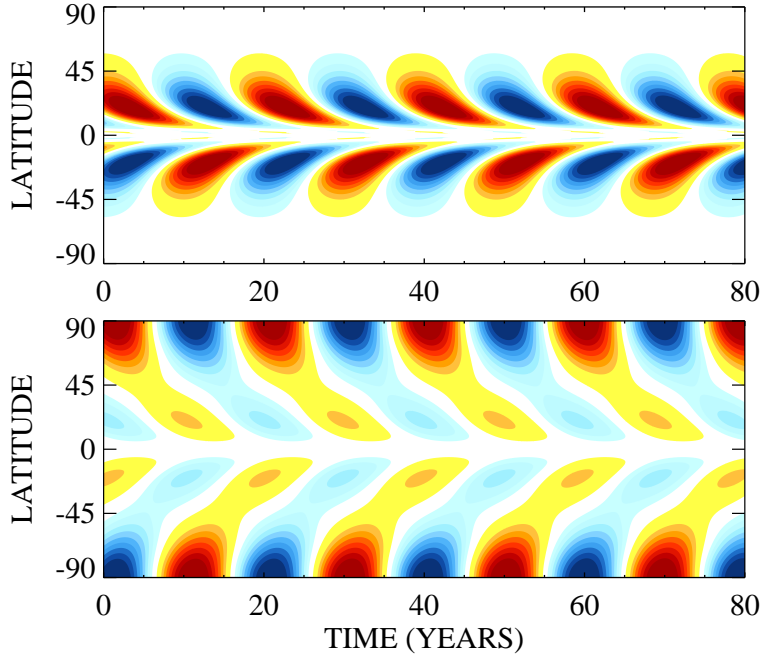}
   \caption{Time-latitude diagram of Model~II for the solar model. Top and bottom panels have
            the same meaning as in Fig\,\ref{f4} but are computed for the $\alpha$-effect of Eq.\,(\ref{9}) and with $P_\mathrm{rot} = P_\odot$.
            }
   \label{f5}
\end{figure}

%%%%%%%%%%%%%%%%%%%%%%%%%%%%%%%%%%%%%%%%%%%%%%%%%%%%%%%%%%%%%%%%%%%%%%%
Model~II with the $\alpha$-effect of Eq.\,(\ref{9}) has not been applied to the Sun before. We therefore consider the solar case first. Figure~\ref{f5} shows the time-latitude diagram computed for $P_\mathrm{rot} = P_\odot$. The polar branch of the poloidal field is lacking but otherwise the diagram is satisfactory. The cycle period in Fig.\,\ref{f5} is about 9.6 years. The cycle period varies little with rotation rate (until a new dynamo regime with a not well defined period onsets for rapid rotation). The reason is the same as for Model~I: the meridional flow is kept constant.

\begin{figure}
   \centering
   \includegraphics[width=10 truecm, angle=0]{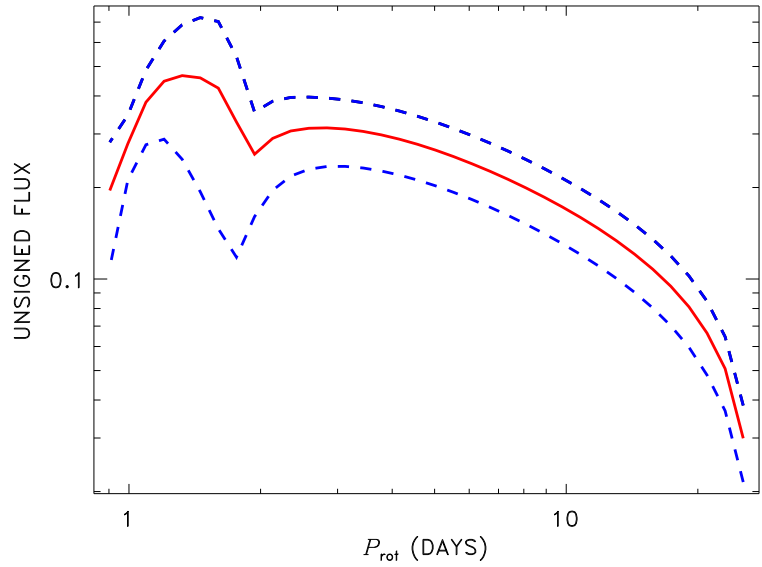}
   \caption{Same as Fig.\,\ref{f3} but for Model~II. The two dashed lines and solid line
            show the maximum and minimum values of $f_\mathrm{m}$ (\ref{12}) in dynamo-cycles and their median value, respectively.
            }
   \label{f6}
\end{figure}

Figure~\ref{f6} shows the dependence of unsigned flux (\ref{12}) on rotation rate. The dependence is not as smooth as in Model~I but generally shows a saturation as well at $P_\mathrm{rot} \lsim 3$\,days. The power index of Eq.\,(\ref{14}) $\delta \simeq 1.3$ (at $P_\mathrm{rot} \simeq 7$\,days) agrees with the computations of \citet{Kea14}.

\begin{figure}
   \centering
   \includegraphics[width=10 truecm, angle=0]{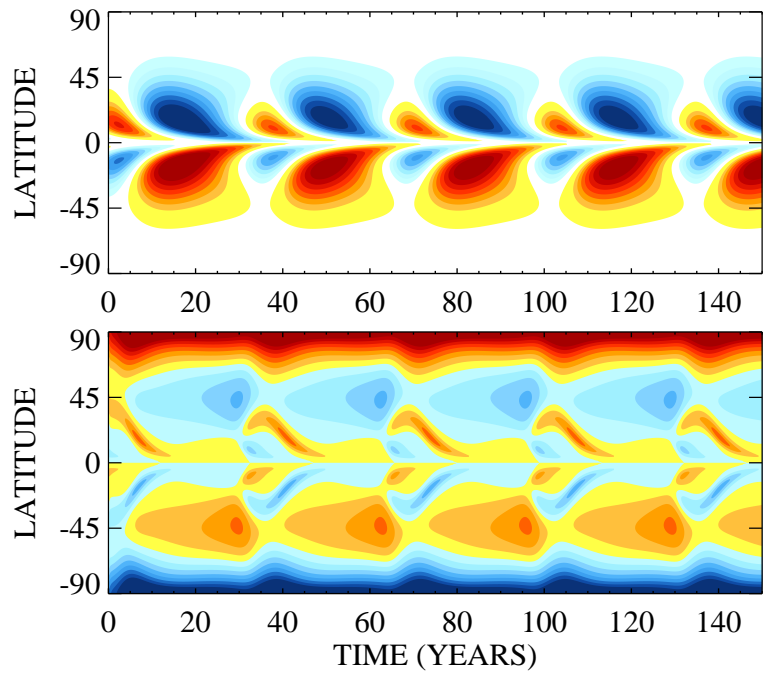}
   \caption{Same as Fig.\,\ref{f5} but for $P_\mathrm{rot} = 1$\,day. The fields of high latitudes oscillate about non-zero mean values.
            }
   \label{f7}
\end{figure}

The kink in the dependence at rotation periods slightly below 2 days signifies a change in the dynamo regime. The new regime is illustrated by the time-latitude diagram of Fig.\,\ref{f7}. This figure shows field reversals at low latitudes but high-latitude fields do not change sign. Polar fields oscillate about a non-zero mean value.

The following interpretation for this behaviour can be suggested. The positive $\alpha$-effect (in the northern hemisphere) of solar-type dynamos generates a poloidal field from the toroidal one, and the new poloidal field has an  opposite sign compared to the old poloidal field from which the toroidal field is wound by the (solar-type) differential rotation. Sign reversals of magnetic fields, therefore, take place and the dynamo is cyclic. If the $\alpha$-effect were negative, a new poloidal field of the same sign as the old one would be generated and steady dynamos could be expected. Figure~\ref{f1} shows sign reversals of the $\alpha$-effect in the northern hemisphere for sufficiently rapid rotation ($P_\mathrm{rot} < 2.2$\,days). The $\alpha$-value remains positive near the equator but changes to negative at high latitudes. When the polar region with negative $\alpha$ is sufficiently broad, a new regime which combines cyclic equatorial with steady polar dynamos sets in.
%%%%%%%%%%%%%%%%%%%%%%%%%%%%%%%%%%%%%%%%%%%%%%%%%%%%%%%%%%%%%%%%%%%%%%%
\section{Conclusion}\label{disc}
%%%%%%%%%%%%%%%%%%%%%%%%%%%%%%%%%%%%%%%%%%%%%%%%%%%%%%%%%%%%%%%%%%%%%%%
The possibility for the saturation of magnetic activity observed for rapidly rotating solar-type stars \citep{V84,Pea03,Wea11} being related to the properties of the BL mechanism for magnetic field generation was discussed. This mechanism naturally saturates when the mean tilt-angle approaches 90$^\circ$. Computations with dynamo models generally support this idea.

Two dynamo models were considered.  The tilt-angle dependence on rotation rate was treated globally in Model~I of Eq.\,(\ref{8}). Model~II of Eq.\,(\ref{9}) employed Joy's law for latitude dependence of the tilt-angles. Both models show saturation of dynamo-generated magnetic flux at high rotation rates.

Model~II is probably more realistic and it demonstrates more variable behaviour. As rotation rate increases, the $\alpha$-effect first saturates, then decreases and can even change sign if the tilt-angle exceeds 180$^\circ$. In Model~II, this happens earlier at higher latitudes (Fig.\,\ref{f1}). This leads to a change in dynamo regime with increasing rotation rate. The new regime emerging at high rotation rates combines field reversals at low latitudes with oscillations about a non-zero mean value without a change of sign in the polar fields (Fig.\,\ref{f7}). Magnetic energy is not sensitive to the field sign and, therefore, shows doubly periodic variations in this dynamo-regime. The double cycles have indeed been observed in solar-type stars with relatively low Rossby numbers \citep{SB99}.

Two versions of the $\alpha$-effect are currently discussed for the Sun: the BL mechanism and the $\alpha$-effect of convective turbulence by \citet{P55}. Evidence for participation of the BL $\alpha$-effect in the solar dynamo is growing \citep{E04,Dea10,OCK13}. Interpretation of dynamo-saturation in rapidly rotating solar-type stars adds one more argument to this line. This is not only because this effect saturates for rapid rotation. The $\alpha$-effect of convective eddies should saturate as well when the eddies are twisted by about $90^\circ$ (the saturation is somehow missed by quasi-linear theories of the $\alpha$-effect). However, the saturation of magnetic activity in the models based on the BL mechanism is found at rotation rates about 20 times the solar value where observations also indicate it occurs in solar-mass stars. A combined action of both effects is possible \citep{Pea14} but the BL mechanism seems to be dominant.

Only the $\alpha$-effect depends on rotation rate in our model. Other parameters can also depend on $P_\mathrm{rot}$ but were kept constant to see the exclusive consequences of $\alpha$-effect saturation. Work combining the model for (variable) stellar differential rotation and meridional flow \citep{KO11_1} with a dynamo model is currently in progress.
%%%%%%%%%%%%%%%%%%%%%%%%%%%%%%%%%%%%%%%%%%%%%%%%%%%%%%%%%%%%%%%%%%%%%%%
\normalem
\begin{acknowledgements}
%%%%%%%%%%%%%%%%%%%%%%%%%%%%%%%%%%%%%%%%%%%%%%%%%%%%%%%%%%%%%%%%%%%%%%%
This work was supported by the Russian Foundation for Basic Research (project no. 13-02-00277).
\end{acknowledgements}
%%%%%%%%%%%%%%%%%%%%%%%%%%%%%%%%%%%%%%%%%%%%%%%%%%%%%%%%%%%%%%%%%%%%%%%
\bibliographystyle{raa}

%%%%%%%%%%%%%%%%%%%%%%%%%%%%%%%%%%%%%%%%%%%%%%%%%%%%%%%%%%%%%%%%%%%%%%%
\end{document}